\pdfoutput=1 

\documentclass[12pt]{article}

\usepackage{scicite}

\usepackage{times}

\usepackage{amsmath}
\usepackage{amsfonts}
\usepackage{amssymb}
\usepackage{graphicx}

\usepackage{float}
\usepackage{bm}

\usepackage[font=doublespacing]{caption}
\usepackage[
singlelinecheck=false 
]{caption}
\usepackage[figurename=Fig.]{caption}

\newcommand{\subfigimg}[3][,]{%
  \setbox1=\hbox{\includegraphics[#1]{#3}}
  \leavevmode\rlap{\usebox1}
  \rlap{\raisebox{\dimexpr\ht1-0\baselineskip}{\textsf{#2}}}
}

\usepackage{tabularray}
\usepackage{pdflscape}
\usepackage{soul}
\usepackage{xcolor}
\sethlcolor{yellow}

\title{
    \textbf{Title:}
    Precursor recommendation for inorganic \\
    synthesis by machine learning materials similarity \\
    from scientific literature
}

\author
{
    \textbf{Authors:} Tanjin He${}^{1,2}$, 
    Haoyan Huo${}^{1,2}$, 
    Christopher J. Bartel${}^{1,2,3}$, \\
    Zheren Wang${}^{1,2}$, 
    Kevin Cruse${}^{1,2}$, 
    Gerbrand Ceder${}^{1,2\ast}$ \\
}

\date{}

\begin{document} 


\baselineskip=24pt


\maketitle



\noindent \textbf{Affiliations}

\noindent ${}^{1}$Department of Materials Science and Engineering, University of California, Berkeley, CA 94720, USA

\noindent ${}^{2}$Materials Sciences Division, Lawrence Berkeley National Laboratory, Berkeley, CA 94720, USA

\noindent ${}^{3}$Department of Chemical Engineering and Materials Science, University of Minnesota, Minneapolis, MN 55455, USA

\noindent $^\ast$Correspondence to: gceder@berkeley.edu

\newpage

\noindent \textbf{Abstract}  

Synthesis prediction is a key accelerator for the rapid design of advanced materials. 
However, determining synthesis variables such as the choice of precursor materials is challenging for inorganic materials because the sequence of reactions during heating is not well understood. 
In this work, we use a knowledge base of 29,900 solid-state synthesis recipes, text-mined from the scientific literature, to automatically learn which precursors to recommend for the synthesis of a novel target material. 
The data-driven approach learns chemical similarity of materials and refers the synthesis of a new target to precedent synthesis procedures of similar materials, mimicking human synthesis design.  
When proposing five precursor sets for each of 2,654 unseen test target materials, the recommendation strategy achieves a success rate of at least 82\%.  
Our approach captures decades of heuristic synthesis data in a mathematical form, making it accessible for use in recommendation engines and autonomous laboratories.  

\bigskip

\noindent \textbf{Short title} 

AI learning inorganic synthesis from literature

\bigskip

\noindent \textbf{Teaser} 

Decades of heuristic data from the literature are automatically captured for guiding successful synthesis of inorganic materials.


\newpage

\section*{MAIN TEXT}

\section*{Introduction}

Predictive synthesis is a grand challenge that would accelerate the discovery of advanced inorganic materials \cite{hemminger2015challenges}. 
The complexity of synthesis mainly originates from the interactions of many design variables, including the diversity of precursor candidates for each element in the target material (oxides, hydroxides, carbonates, etc.), the experimental conditions (temperature, atmosphere, etc.), and the chronological organization of operations (mixing, firing, reducing, etc.).
Properly selecting the combination of experimental variables is crucial and demanding for successful synthesis \cite{miura2021observing,bianchini2020interplay,jiang2017situ}.
Here, we focus on the rational design of precursor combinations for solid-state synthesis, a widely used approach to create inorganic materials.

Because of the lack of a general theory for how phases evolve during heating, synthesis design is mostly driven by heuristics and basic chemical insights.
Unlike the success of retrosynthesis and automated design for organic materials based on the conservation and transformation of functional groups \cite{corey1988robert,stein1993turning,segler2018planning}, the mechanisms underlying inorganic solid-state synthesis are not well understood \cite{chamorro2018progress,kohlmann2019looking,stein1993turning,schafer1971preparative}.
Here, we define a recipe to be any structured information about a target material, including the precursors, operations, conditions, and other experimental details. 
Experimental researchers usually approach a new inorganic synthesis by manually looking up similar materials in the literature and repurposing precedent recipes for a novel material.
However, deciding what materials are similar and thus where to look is often driven by intuition and limited by individuals' personal experience in specific chemical spaces, hindering the ability to rapidly design syntheses for new chemistries.
With the emergence of large-scale materials synthesis datasets from text-mining efforts \cite{kononova2019text,kim2017machine,swain2016chemdataextractor,hiszpanski2020nanomaterial}, it is becoming possible to statistically learn the similarity of materials and the correlation of their synthesis variables in a more systematic and quantitative fashion, and provide such tools as a guide to scientists when approaching the synthesis of novel compounds.

Several studies have demonstrated the promise of building general models for the predictive synthesis of inorganic materials.
Aykol et al. \cite{aykol2021rational} and McDermott et al. \cite{mcdermott2021graph} proposed heuristic models to rank the favorability of synthesis reactions or pathways based on thermodynamic metrics such as the reaction energy, nucleation barrier, and the number of competing phases. 
Kim et al. \cite{kim2020inorganic} utilized the stochasticity of a conditional variational autoencoder model to generate various samples of synthesis actions and precursors for the target material.
Huo et al. \cite{huo2022machine} predicted synthesis conditions using large solid-state synthesis datasets text-mined from scientific journal articles.
An interesting yet unexplored angle is to machine learn how the precursors of different target materials are shared and varied to enable the recommendation of multiple synthesis recipes with some ranked potential of success.
In addition, extending the assessment from specific case studies to a large test set is also valuable for the development and improvement of predictive synthesis models.

\begin{figure}[H]
	\centering
	\includegraphics[width=1.0\linewidth]{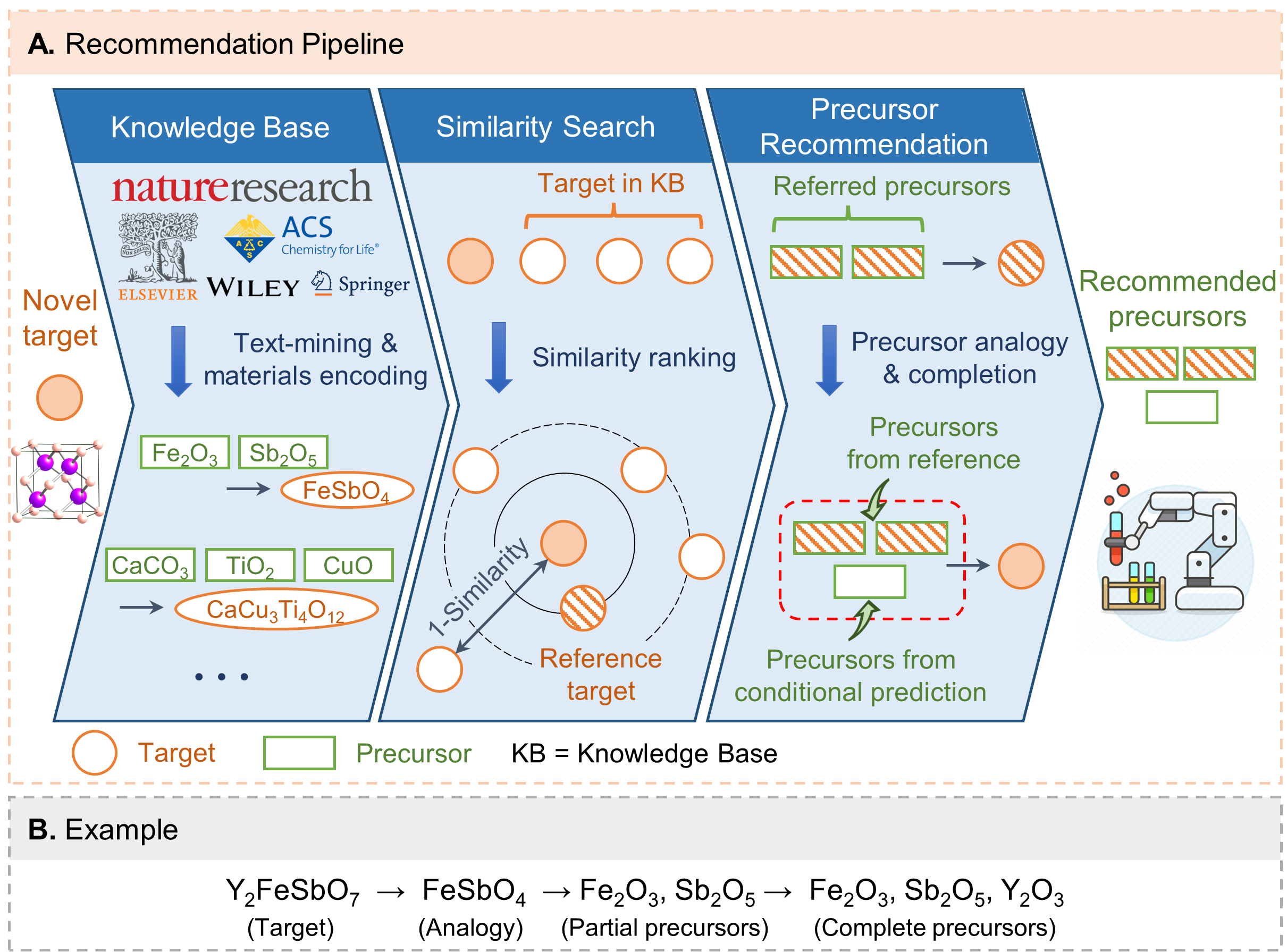}
	\caption{
		\textbf{Precursor recommendation strategy.} 
		\textbf{(A)} Pipeline for precursor recommendation consisting of three steps: (1) digitize target materials in the synthesis knowledge base text-mined from scientific literature, (2) rank target materials in the knowledge base according to the similarity to the novel target, and (3) recommend precursors based on analogy to the most similar target. 
		\textbf{(B)} An example of precursor recommendation for $ \operatorname{Y_2FeSbO_7} $ by referring to the synthesis of $ \operatorname{FeSbO_4} $.
	}
	\label{fig:pipeline}
\end{figure}

We propose a precursor recommendation strategy (Fig. \ref{fig:pipeline}) based on machine-learned similarity of materials to automate the literature-based approach used by experimental researchers.
Inspired by natural language processing (NLP) models \cite{mikolov2013efficient,mikolov2013distributed,devlin2018bert}, we designed an encoding neural network to learn the vectorized representation of a material based on its corresponding precursors for the quantification of materials similarity. 
Assuming that the target material can be synthesized using an experimental design adapted from a similar material, synthesis variables such as precursors, operations, and conditions can be proposed and ranked by querying the knowledge base of previously synthesized materials.
In this work, we applied the recommendation strategy to predict precursors for 2,654 test target materials in a historical validation.
Learning from a knowledge base of 29,900 synthesis reactions text-mined from the scientific literature, we demonstrate that the algorithm can acquire chemical knowledge on materials similarity via self-supervised learning, and make promising decisions on precursor selection.
Our quantitative recommendation pipeline captures how experimental researchers learn synthesis from the literature and enables rational and rapid precursor selection for new inorganic materials.
It also provides meaningful initial solutions in the active learning and decision-making process for autonomous synthesis.

\section*{Results}

We begin with statistical insights from solid-state synthesis experiments reported in 24,304 papers \cite{kononova2019text} to better understand the problem of precursor selection (Section ``Problem of precursor selection'').
Because a universal model for solid-state synthesis has not yet been established, we use a data-driven method to recommend potential precursor sets for the given target material (Fig. \ref{fig:pipeline}). 
The recommendation pipeline consists of three steps: (i) an encoding model to digitize the target material as well as known materials in the knowledge base (Section ``Materials encoding for precursor selection''), (ii) similarity query based on the materials encoding to identify a reference material that is most similar to the target (Section ``Similarity of target materials''), and (iii) recipe completion to (a) compile the precursors referred from the reference material and (b) add any possibly missed precursors if element conservation is not achieved using conditional predictions based on referred precursors (Section 
``Recommendation of precursor materials'').

\subsection*{Problem of precursor selection} \label{sec:problem_description} 

In the solid-state synthesis of inorganic materials, precursor selection plays a crucial role in governing the synthesis pathway by yielding intermediates that may lead to the desired material or alternative phases 
\cite{miura2021observing,bianchini2020interplay,jiang2017situ}. 

\begin{figure}[H]
    \centering
    \begin{tabular}{@{}p{0.5\linewidth}@{}p{0.5\linewidth}@{}}
        \subfigimg[width=\linewidth]{\textbf{A}}{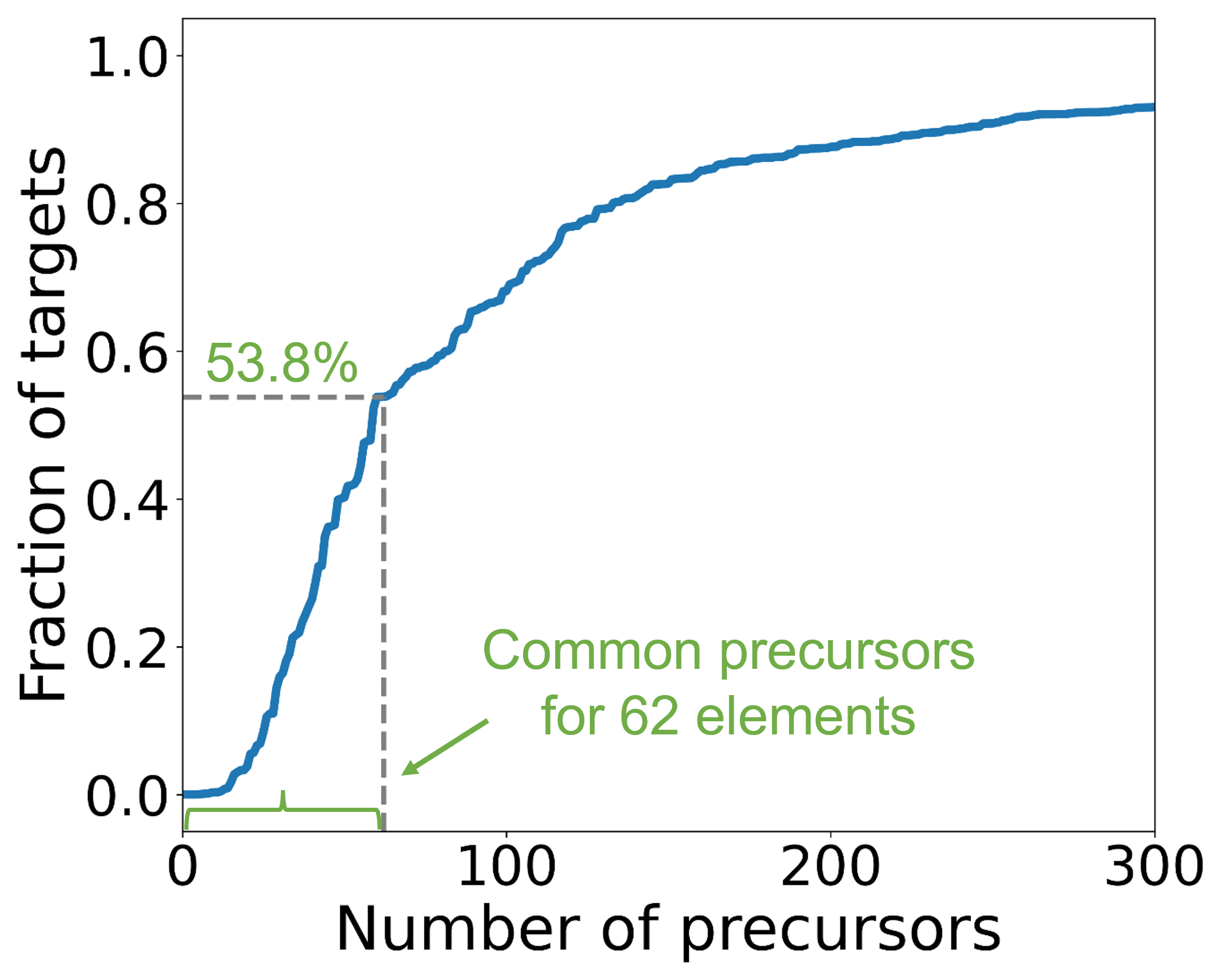} &
        \subfigimg[width=\linewidth]{\textbf{B}}{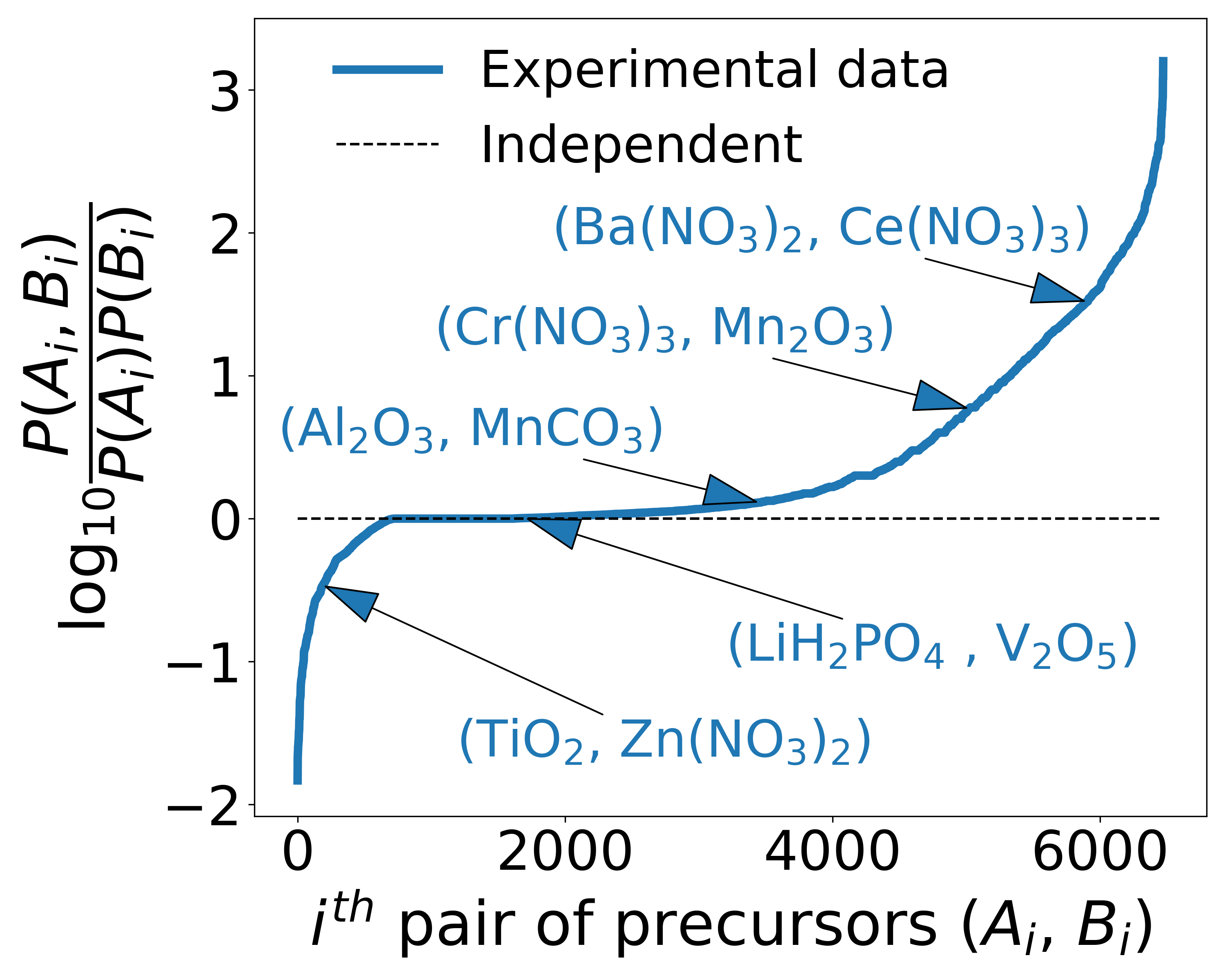}
    \end{tabular}
	\caption{
	    \textbf{Usage of precursors in solid-state synthesis.}
		\textbf{(A)} Fraction of targets that can be synthesized with limited number of available precursors. 
		The precursors are ordered by relative frequency per metal/metalloid element.
        Precursors for 62 elements are considered.
        A target is included if at least one reported reaction for that target was performed with the available precursors. 
 	\textbf{(B)} Pairwise dependency of precursors $A_i$ and $B_i$ characterized by $ \frac{P(A_i,B_i)}{P(A_i)P(B_i)} $. 
    	Probability is estimated from the frequency of occurrence in the solid-state synthesis dataset.
    	The value of $ \log_{10}\frac{P(A_i,B_i)}{P(A_i)P(B_i)} $ is zero when $A_i$ and $B_i$ are independent, positive when $A_i$ and $B_i$ tends to be used in the same experiment more frequently than $ P(A_i)P(B_i) $, negative otherwise.  
	}
	\label{fig:precursor_statistics}
\end{figure}

For each metal/metalloid element, one precursor is often used predominantly over all others, which we denote as the common precursor \cite{he2020similarity}. 
However, in a solid-state synthesis dataset of 33,343 experimental recipes extracted from 24,304 materials science papers \cite{kononova2019text}, we find that approximately half of the target materials were synthesized using at least one uncommon precursor.
Fig. \ref{fig:precursor_statistics}A presents the fraction of targets in the text-mined dataset \cite{kononova2019text} that can be achieved as one increases the number of available precursors. 
The precursors on the x-axis are ordered by the relative frequency with which they are used to bring a specific element into a synthesis target.
Uncommon precursors may be used for a variety of reasons including synthetic constraints (e.g., temperature and time), purity, morphology, and anthropogenic factors \cite{he2020similarity,jia2019anthropogenic,miura2021observing}.

In addition, a probability analysis of the text-mined dataset indicates that precursors for different chemical elements are not randomly combined.
The joint probability to select a specific precursor pair ($A_i$, $B_i$) can be compared to the marginal probability to select $A_i$ for element $Ele_a$ and $B_i$ for $Ele_b$. 
If the choices of $A_i$ and $B_i$ are independent, the joint probability should equal the product of the marginal probabilities, namely, $P(A_i, B_i) = P(A_i)P(B_i)$.
However, inspection of 6,472 pairs of precursors from our text-mined dataset (Fig. \ref{fig:precursor_statistics}B) reveals that many show a strong dependency on each other (i.e., $P(A_i, B_i)$ deviating significantly from $P(A_i)P(B_i)$). 
A well-known example is that nitrates such as $ \operatorname{Ba(NO_3)_2} $ and $ \operatorname{Ce(NO_3)_3} $ tend to be used together, likely because of their solubility and applicability for solution processing (e.g. slurry preparation).
Unfortunately, these decisions regarding dependencies of precursors are usually empirical and hard to standardize.
Machine learning is a possible solution to ingest the heuristics that underlie such selections.

\subsection*{Materials encoding for precursor selection} \label{sec:encoding_model} 

Our precursor recommendation model for the synthesis of a novel target will mimic the human approach of trying to identify similar target materials for which successful synthesis reactions are known.
To find similar materials, digital processing requires an encoding model that transforms any arbitrary inorganic material into a numerical vector.
For organic synthesis, structural fingerprinting such as Morgan2Feat \cite{rogers2010extended} is a good choice \cite{coley2017computer} because it is natural to track the conservation and change of functional groups in organic reactions, but the concept of functional groups is not applicable to inorganic synthesis.
Chemical formulas of inorganic solids have been represented using a variety of approaches (e.g., Magpie \cite{ward2016general,ward2018matminer}, Roost \cite{goodall2020predicting}, CrabNet \cite{wang2021compositionally}). 
However, these representations are typically used as inputs to predict thermodynamic or electronic properties of materials.
Here, we attempt to directly incorporate synthesis information into the representation of a material with arbitrary composition.
Local text-based encodings such as Word2Vec 
\cite{tshitoyan2019unsupervised,pei2023toward} and FastText \cite{kim2020inorganic} are able to capture contextual information from the materials science literature, of which synthesis information is a part; 
however, they are not applicable to unseen materials when the materials text (sub)strings are not in the vocabulary or when the materials are not in the predefined composition space.
For example, Pei et al. \cite{pei2023toward} computed the similarity of high-entropy alloys as the average similarity of element strings by assuming the elements are present in equal proportions in the material (e.g., CoCrFeNiV). 
However, this approach is not applicable to unseen materials different from such composition template, and consequently would not be practical in our work on synthesis of diverse inorganic materials.
Substitution modeling can evaluate similarity of precursors by assessing the viability of substituting one precursor with another while retaining the same target, but it cannot be used to identify analogues for new target materials \cite{he2020similarity}.
In this work, we propose a synthesis context-based encoding model utilizing the idea that target materials produced with similar synthesis variables are similar.

\begin{figure}[H]
	\centering
	\includegraphics[width=1.0\linewidth]{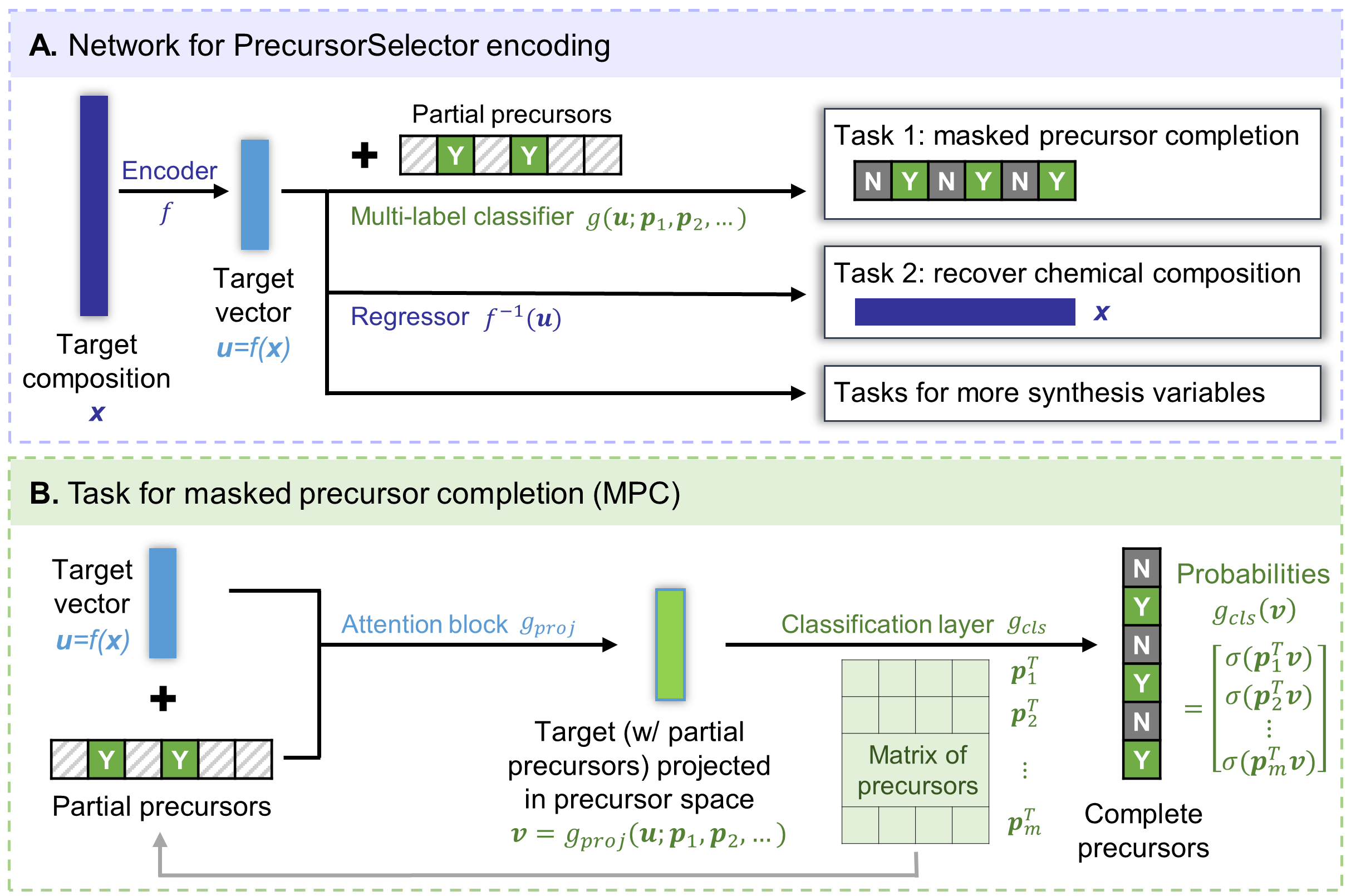}
	\caption{
		\textbf{Representation learning to encode precursor information for target materials.}
		\textbf{(A)} Multi-task network structure to encode the target material in the upstream and to predict the complete precursor set, chemical composition, and more synthesis variables in the downstream.
		$\bm{x}$ and $\bm{u}$ represent the composition and encoded vector of the target material, respectively.
		$\bm{p}_i$ represents the $i^{\operatorname{th}}$ precursor in a predefined ordered precursor list. 
		Dense layers are used in each layer unless specified differently.
		\textbf{(B)} Submodel of multi-label classification for the masked precursor completion (MPC) task. 
		Part of the precursors are randomly masked; the remaining precursors (marked as ``Y'') are used as a condition to predict the probabilities of other precursors for the target material.
		The probabilities corresponding to the complete precursors (marked as ``Y'') are expected to be higher than that of unused precursors (marked as ``N'').
		The attention block $g_{proj}$ \cite{vaswani2017attention} is used to aggregate the target vector and conditional precursors.
		The final classification layer $g_{cls}$ and the embedding matrix for conditional precursors share the same weights. 
		$\sigma$ represents the sigmoid function.
	}
	\label{fig:NN_framework}
\end{figure}

Analogous to how language models \cite{mikolov2013efficient,mikolov2013distributed,devlin2018bert} pre-train word representations by predicting context for each word, we use a self-supervised representation learning model to encode arbitrary materials by predicting precursors for each target material, which we refer to as PrecursorSelector encoding (Fig. \ref{fig:NN_framework}A).
The upstream part is an encoder where properties of the target material are projected into a latent space as the encoded vector representation.
In principle, any intrinsic materials property could be included at this step.
Here, we use only composition for simplification.
The downstream part consists of multiple tasks where the encoded vector is used as the input to predict different variables related to precursor selection.
Here, we use a masked precursor completion (MPC) task (Fig. \ref{fig:NN_framework}B) to capture (i) the correlation between the target and precursors and (ii) the dependency between different precursors in the same experiment.
For each target material and corresponding precursors in the training set, we randomly mask part of the precursors and use the remaining precursors as a condition to predict the complete precursor set.
We also add a task of reconstructing the chemical composition to conserve the compositional information of the target material. 
The downstream task part is designed to be extensible; other synthesis variables such as operations and conditions can be incorporated by adding corresponding prediction tasks in a similar fashion.
By training the entire neural network, the encoded vectors for target materials with similar precursors are automatically dragged closer to each other in the latent space because that reduces the overall prediction error.
This PrecursorSelector encoding thus takes the correlation induced by precursor selection and serves as a useful metric to measure similarity of target materials in syntheses.

To demonstrate that the neural network is able to learn precursor information, we present the results of the MPC task (Fig. \ref{fig:NN_framework}B) for $ \operatorname{LaAlO_3} $ as an example (Table \ref{table:conditional_prediction}).
$ \operatorname{LaAlO_3} $ is a ternary material that normally requires two precursors (one to deliver each cation, La and Al).
In this test, we masked one precursor and asked the model to predict the complete precursor set. 
For the same target conditioned with different partial precursors, the predicted probabilities of precursors strongly depend on the given precursor and agree with some rules of thumb for precursor selection.
When the partial precursors are oxides such as $ \operatorname{La_2O_3} $ or $ \operatorname{Al_2O_3} $, the most probable precursors are predicted to be oxides for the other element, i.e., $ \operatorname{Al_2O_3} $ for $ \operatorname{La_2O_3} $ and $ \operatorname{La_2O_3} $ for $ \operatorname{Al_2O_3} $ \cite{mao2011investigation}. 
When the partial precursors are nitrates such as $ \operatorname{La(NO_3)_3} $ or $ \operatorname{Al(NO_3)_3} $, nitrates for the other element are prompted with higher probabilities, i.e., $ \operatorname{Al(NO_3)_3} $ for $ \operatorname{La(NO_3)_3} $ and $ \operatorname{La(NO_3)_3} $ for $ \operatorname{Al(NO_3)_3} $ \cite{mendoza2012molten}. 
If both precursors are masked, oxides rank first in the prediction because the common precursors for elements La and Al are $ \operatorname{La_2O_3} $ and $ \operatorname{Al_2O_3} $, respectively.
The simple successful prediction shows our PrecursorSelector encoding model is able to learn the correlation between the target and precursors in different contexts of synthesis without explicit input of chemical rules about synthesis.
In addition, the use of different precursors suggests various synthetic routes may lead to the same target material. 
When a practical preference for a particular route exists, the framework we introduce in this work can be extended to include more constraints, such as synthesis type, temperature, morphology, particle size, and cost of precursors, by learning from pertinent datasets \cite{wang2022dataset,cruse2022text,jia2019anthropogenic}.

\begin{table}[H]
	\caption{
    	\textbf{MPC conditioned on different partial precursors for the same target material $ \operatorname{LaAlO_3} $.} 
    	The predicted complete precursors are the ones with the highest probabilities (bold).
        The term ``N/A'' denotes the absence of partial precursors, i.e., all precursors are masked in the MPC task.
	}
	\centering
	\setstretch{1.5}
     \begin{tblr}{
        colspec={ccccccc},
    	rowsep=0pt,
    	hline{1,Z}={2pt,solid},
        hline{2-Y}={1pt,solid},
        hspan=minimal,
    }
	\SetCell[r=2]{c} { Partial precursors \\ (condition) }
	& \SetCell[c=6]{c} Probability to use different precursors (output)  \\
	& $ \operatorname{La_2O_3} $ 
	& $ \operatorname{Al_2O_3} $ 
	& $ \operatorname{La(NO_3)_3} $ 
	& $ \operatorname{Al(NO_3)_3} $ 
	& $ \operatorname{La_2(CO_3)_3} $  
	& $ \operatorname{Al(OH)_3} $    \\
	$ \operatorname{La_2O_3} $    
	& \textbf{0.75} & \textbf{0.71} & 0.58          & 0.57          & 0.57 & 0.57 \\
	$ \operatorname{Al_2O_3} $    
	& \textbf{0.72} & \textbf{0.73} & 0.58          & 0.57          & 0.58 & 0.56 \\
	$ \operatorname{La(NO_3)_3} $ 
	& 0.60          & 0.59          & \textbf{0.64} & \textbf{0.63} & 0.61 & 0.61 \\
	$ \operatorname{Al(NO_3)_3} $ 
	& 0.62          & 0.58          & \textbf{0.65} & \textbf{0.65} & 0.62 & 0.60 \\
	N/A      
	& \textbf{0.70} & \textbf{0.69} & 0.59          & 0.58          & 0.59 & 0.59 \\
	\end{tblr}
	\label{table:conditional_prediction}
\end{table}

\subsection*{Similarity of target materials} \label{sec:target_similarity}

Similarity establishes a link between a novel material to synthesize and the known materials in the knowledge base because it is reasonable to assume similar target materials share similar synthesis variables in experiments.
Although the understanding of similarity is generally based on heuristics, the PrecursorSelector encoding introduced in Section ``Materials encoding for precursor selection'' provides a meaningful representation for quantified similarity analysis. 
Dedicated to precursor prediction in this study, we define the similarity of two target materials as the similarity of the precursors used in their respective syntheses. 
Although precursors for a new target material are not known in advance, the PrecursorSelector encoding serves as a proxy reflecting the potential precursors to use. 
In that latent space, we can take the cosine similarity \cite{mikolov2013efficient,mikolov2013distributed,tshitoyan2019unsupervised} of the PrecursorSelector encoding as a measure of the similarity (Sim) of two target materials $ \bm{x}_1 $ and $ \bm{x}_2 $:
\begin{equation} \label{eq:similarity_encoding}
	\begin{aligned}
		\operatorname{Sim}(\bm{x}_1, \bm{x}_2) \sim \cos(f(\bm{x}_1), f(\bm{x}_2) ),
	\end{aligned}
\end{equation}
where $f$ is the encoder part of the PrecursorSelector model transforming the composition of the target material $\bm{x}$ into the encoded target vector (Fig. \ref{fig:NN_framework}A).

To demonstrate that the similarity estimated from PrecursorSelector encoding is reasonable, we show typical materials with different levels of similarity to an example target material $ \operatorname{NaZr_2(PO_4)_3} $ (Table \ref{table:similarity_ranking}).
The most similar materials are the ones with the same elements such as Zr-containing phosphates and other sodium super ionic conductor (NASICON) materials. 
The similarity decreases slightly as additional elements are introduced (e.g., $ \operatorname{Na_3Zr_{1.9}Ti_{0.1}Si_2PO_{12}} $) or when one element is substituted (e.g., $ \operatorname{LiZr_2(PO_4)_3} $).
When the phosphate groups are replaced with another anion, the similarity decreases further, with oxides having generally mild similarity to the phosphate $ \operatorname{NaZr_2(PO_4)_3} $.
The similarity decreases even further for compounds with no anion (e.g., intermetallics) and for non-oxygen anions (e.g., chalcogenides).
This finding agrees with our experimental experience that when seeking a reference material, researchers will usually refer to compositions in the same chemical system or to cases where some elements are substituted.
It is also worth noting that our quantitative similarity is purely a data-driven abstraction from the literature and uses no externally chemical knowledge.

\begin{table}[H]
	\caption{
 \textbf{Different levels of similarity between $ \operatorname{NaZr_2(PO_4)_3} $ and materials in the knowledge base.}
    }
	\centering
	\setstretch{1.5}
    \begin{tblr}{
        colspec={cccc},
    	rowsep=0pt,
    	hline{1,Z}={2pt,solid},
        hline{2-Y}={1pt,solid},
        hspan=minimal,
    }
		Target & Similarity & Target & Similarity   \\
		$ \operatorname{Zr_3(PO_4)_4} $                       & 0.946  & $ \operatorname{Li_{1.8}ZrO_3} $            & 0.701   \\
		$ \operatorname{Na_3Zr_2Si_2PO_{12}} $                & 0.929  & $ \operatorname{NaNbO_3} $                  & 0.600   \\
		$ \operatorname{Na_3Zr_{1.8}Ge_{0.2}Si_2PO_{12}} $    & 0.921  & $ \operatorname{Li_2Mg_2(MoO_4)_3} $ 	     & 0.500   \\
		$ \operatorname{Na_3Ca_{0.1}Zr_{1.9}Si_2PO_{11.9}} $  & 0.908  & $ \operatorname{Sr_2Ce_2Ti_5O_{16}} $       & 0.400   \\
		$ \operatorname{Na_3Zr_{1.9}Ti_{0.1}Si_2PO_{12}} $    & 0.900  & $ \operatorname{Ga_{0.75}Al_{0.25}FeO_3} $  & 0.300   \\
		$ \operatorname{LiZr_2(PO_4)_3} $                     & 0.896  & $ \operatorname{Cu_2Te} $                   & 0.200   \\
		$ \operatorname{NaLa(PO_3)_4} $                       & 0.874  & $ \operatorname{Ni_{60}Fe_{30}Mn_{10}} $    & 0.100   \\
		$ \operatorname{Sr_{0.125}Ca_{0.375}Zr_2(PO_4)_3} $   & 0.852  & $ \operatorname{AgCrSe_2} $ 	             & 0.000   \\
		$ \operatorname{Na_5Cu_2(PO_4)_3} $                   & 0.830  & $ \operatorname{Zn_{0.1}Cd_{0.9}Cr_2S_4} $  & -0.099  \\
		$ \operatorname{LiGe_2(PO_4)_3} $                     & 0.796  & $ \operatorname{Cr_2AlC} $                  & -0.202  \\
	\end{tblr}
	\label{table:similarity_ranking}
\end{table}

\begin{figure}[H]
	\centering
	\includegraphics[width=0.50\linewidth]{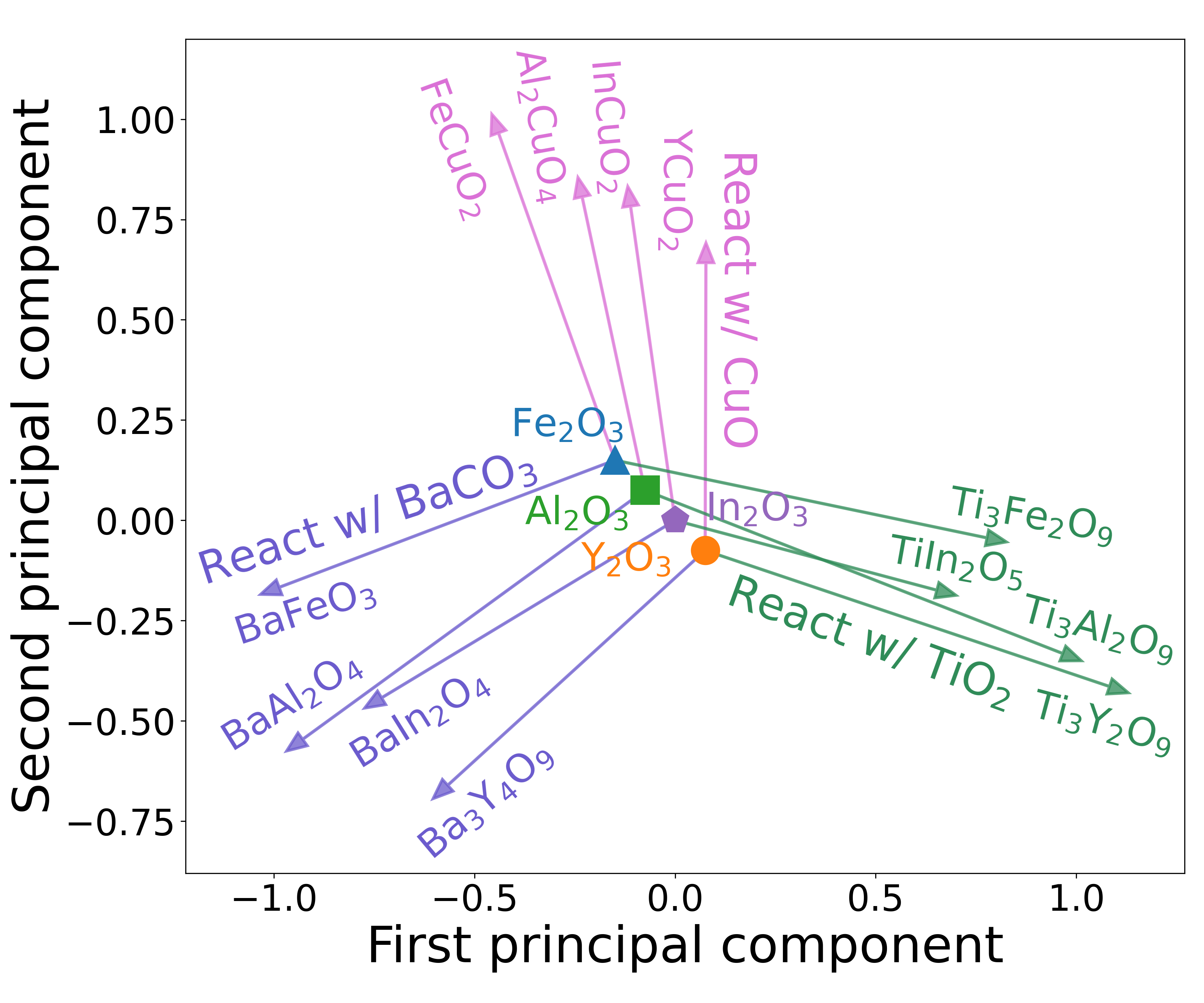}
	\caption{
		\textbf{Relationships between targets and their shared precursors.}
		Four groups of target materials are synthesized each using one shared precursor shown as the original point ($ \operatorname{Y_2O_3} $, $ \operatorname{In_2O_3} $, $ \operatorname{Al_2O_3} $,  or $ \operatorname{Fe_2O_3} $) and one distinct precursor shown as the edge ($ \operatorname{BaCO_3} $, $ \operatorname{CuO} $, or $ \operatorname{TiO_2} $).
		The relationship of ``react with another precursor'' is visualized as the first two principal components of the difference vector between the target and the shared precursor $ g_{proj}(f(\bm{x}))-\bm{p}_i $. 
		The original points corresponding to different precursors $ \bm{p}_i $'s are jittered for clarity. 
	}
	\label{fig:relationship_shift}
\end{figure}

\bigskip
To better understand the similarity, we conducted a relationship analysis \cite{mikolov2013efficient,mikolov2013distributed,tshitoyan2019unsupervised} by visualizing 
four groups of target materials synthesized using one shared precursor and one distinct precursor (Fig. \ref{fig:relationship_shift}).
For example, the syntheses of $ \operatorname{YCuO_2} $, $ \operatorname{Ba_3Y_4O_9} $, and $ \operatorname{Ti_3Y_2O_9} $ share $ \operatorname{Y_2O_3} $ as a precursor and separately use $ \operatorname{CuO} $, $ \operatorname{BaCO_3} $, and $ \operatorname{TiO_2} $.
The three other groups share the precursors $ \operatorname{In_2O_3} $, $ \operatorname{Al_2O_3} $, and $ \operatorname{Fe_2O_3} $, respectively.
To separate the effect of the precursor variation, we align the original points of the target vectors by first projecting each target vector to the same vector space as the precursors and then subtracting the vector of the shared precursor, providing a difference vector showing the relationship between the target material and the shared precursor (more details in Section ``Representation learning for similarity of materials'').
Next, we plot the top two principal components \cite{pearson1901liii} of these difference vectors in a two-dimensional plane.
The difference vectors are automatically separated into three clusters according to the precursor variate, representing three types of relationships, ``react with $ \operatorname{BaCO_3} $'', ``react with $ \operatorname{CuO} $'', and ``react with $ \operatorname{TiO_2} $'', respectively. 
For example, $ \operatorname{Ba_3Y_4O_9} $ is to $ \operatorname{Y_2O_3} $ as $ \operatorname{BaAl_2O_4} $ is to $ \operatorname{Al_2O_3} $ (i.e., $ \operatorname{Ba_3Y_4O_9} - \operatorname{Y_2O_3} \approx \operatorname{BaAl_2O_4} - \operatorname{Al_2O_3} $) because both syntheses use $ \operatorname{BaCO_3} $.
The consistency between this automatic clustering and the chemical intuition again affirms the efficacy of using PrecursorSelector encoding as a similarity metric.

\subsection*{Recommendation of precursor materials} \label{sec:precursor_recommendation}

With the capability of measuring similarity, a natural solution to precursor selection is to replicate the literature-based approach used by experimental researchers.
Given a novel material to synthesize, we initialize our recommendation by first proposing a recipe consisting of common precursors for each metal/metalloid element in the target material because this might be the first attempt in a lab.
Then, we encode the novel target material and known target materials in the knowledge base using PrecursorSelector encoding model from Section ``Materials encoding for precursor selection'' and calculate the similarity between the novel target and each known material with Eq. \ref{eq:similarity_encoding}.
We rank known materials based on their similarity to the target such that a reference material can be identified that is the most similar to the novel target.
When the precursors used in the synthesis of the reference material cannot cover all elements of the target, we use MPC in Fig. \ref{fig:NN_framework}B to predict the missing precursors.
For example, for $ \operatorname{Y_2FeSbO_7} $ (Fig. \ref{fig:pipeline}B), the most similar material in the knowledge base is $ \operatorname{FeSbO_4} $.
It is reasonable to assume that the precursors $ \operatorname{Fe_2O_3} $ and $ \operatorname{Sb_2O_5} $ used in the synthesis of $ \operatorname{FeSbO_4} $ \cite{zvereva2013new} can also be used to synthesize $ \operatorname{Y_2FeSbO_7} $.
Because the Y source is missing, MPC finds $ \operatorname{Y_2O_3} $ is likely to fit with $ \operatorname{Fe_2O_3} $ and $ \operatorname{Sb_2O_5} $ for the synthesis of $ \operatorname{Y_2FeSbO_7} $, ending up as a complete precursor set ($ \operatorname{Fe_2O_3} $, $ \operatorname{Sb_2O_5} $, and $ \operatorname{Y_2O_3}) $ \cite{luan2011preparation}.
Multiple attempts of recommendation are feasible by moving down the list of known materials ranked to be most similar to the novel target. 

\begin{figure}[H]
	\centering
	\includegraphics[width=0.50\linewidth]{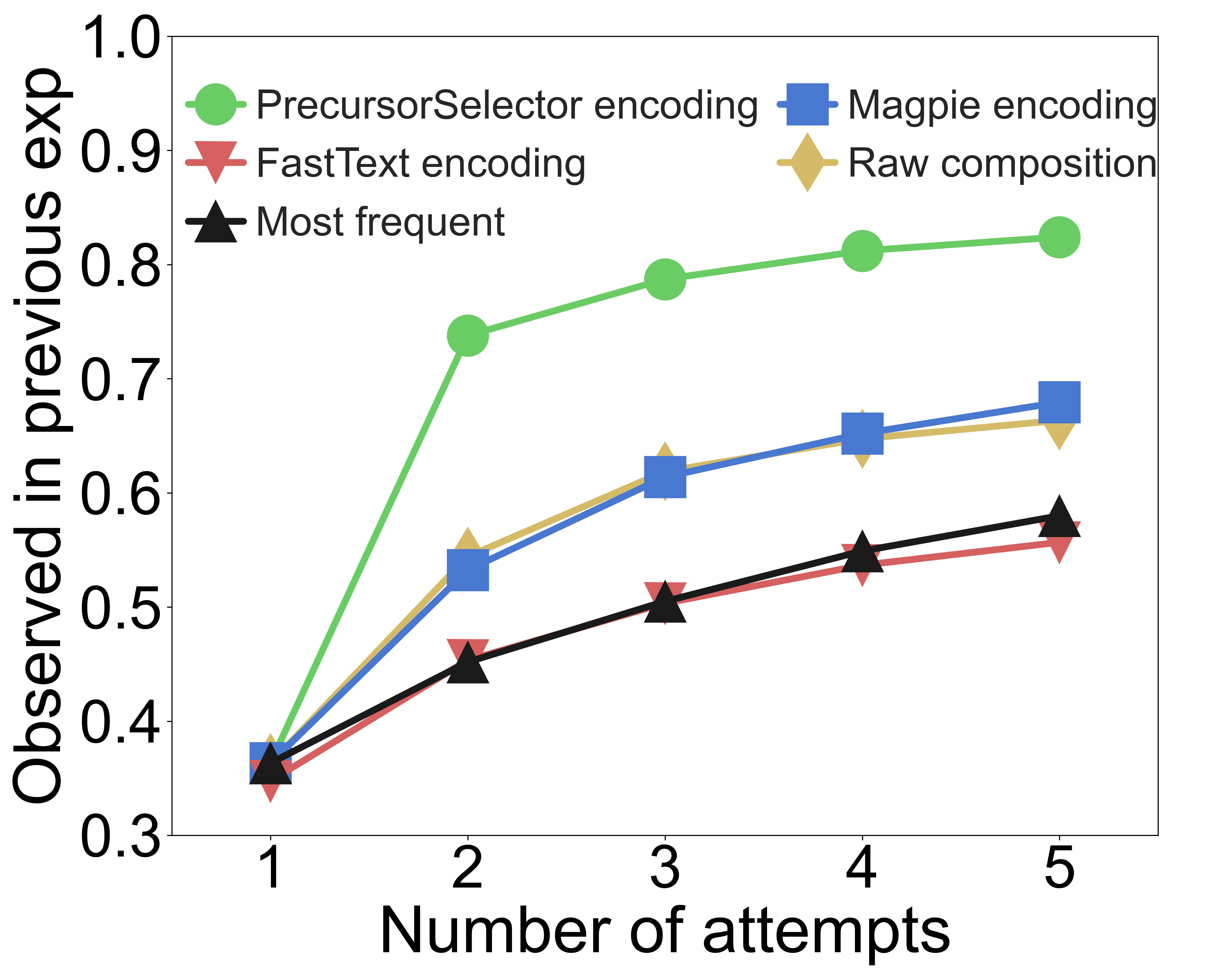}
	\caption{
        \textbf{Performance of various precursor prediction algorithms.}
		For each of the 2,654 test target materials, the algorithm attempts to propose $ n $ ($ 1 \leq n \leq 5 $ as the x-axis) precursor sets. 
		The y-axis shows the success rate that at least one out of the $ n $ proposed precursor set is observed in previous experimental records.
		PrecursorSelector encoding: this work.
		Magpie encoding/FastText encoding/Raw composition: similar recommendation pipeline to this work but using Magpie representation\cite{ward2016general,ward2018matminer}/FastText representation\cite{kim2020inorganic}/the raw material composition.
		Most frequent: select precursors by frequency.
	}
	\label{fig:precursors_multi_attempts}
\end{figure}

To evaluate our recommendation pipeline, we conduct a validation (Fig. \ref{fig:precursors_multi_attempts}) using the 33,343 synthesis recipes text-mined from the scientific literature.
Using the knowledge base of 24,034 materials reported by the year 2014, we predict precursors for 2,654 test target materials newly reported from 2017 to 2020  (more details in Section ``Data preparation''). 
Because multiple precursors exist for each element, the number of possible precursor combinations increases combinatorially with the number of elements present in the target material. 
A good precursor prediction algorithm is anticipated to select from hundreds of possible precursor combinations those that have a higher probability of success.
For each test material, we attempt to propose five different precursor sets.
For each attempt, we calculate the percentage of test materials being successfully synthesized, where success means at least one set of proposed precursors has been observed in previous experiments.
The similarity-based reference already increases the success rate to 73\% at the second attempt. 
The first guess is set to default to the most common precursors which leads to 36\% success rate.
Within five attempts, the success rate of our recommendation pipeline using PrecursorSelector encoding is 82\%, comparable to the performance of recommendations for organic synthesis \cite{coley2017computer}. 
We note that as defined here, ``success'' will be underestimated since some suggested precursor sets may actually lead to successful target synthesis even though they may not have been tried (and therefore do not appear in the data).

We also establish a baseline model (``Most frequent" in Fig. \ref{fig:precursors_multi_attempts}) that ranks precursor sets based on the product of frequencies with which different precursors are used in the literature (more details in Section ``Baseline models'').
This baseline simulates the typical early stage of the trial-and-error process where researchers grid-search different combinations of precursors matching elements present in the target material without the knowledge of dependency of precursors (Fig. \ref{fig:precursor_statistics}B).
The success rate of this baseline is 58\% within five attempts.
Our recommendation pipeline performs better because the dependency of precursors is more easily captured when the combination of precursors is sourced from a previously used successful recipe for a similar target. 
Through in-situ diffraction of synthesis \cite{miura2021observing,bianchini2020interplay,jiang2017situ}, it is now better understood that some precursor sets do not lead to the target material because they form intermediate phases which have consumed much of the overall reaction energy, thereby leaving a low driving force to form the target. 
It is likely that our literature informed precursor prediction approach implicitly captures some of this reactivity and pathway information, resulting in a higher prediction power than random selection or selection based on how common a precursor is.

In addition, we compare with three other baseline models (``Magpie encoding'', ``FastText encoding'', and ``Raw composition'' in Fig. \ref{fig:precursors_multi_attempts}) using the same recommendation strategy but different encoding methods (more details in Section ``Baseline models''). 
Magpie encoding \cite{ward2016general,ward2018matminer} is a set of attributes computed using the fraction of elements in a material, including stoichiometric attributes, elemental property statistics, electronic structure attributes, and ionic compound attributes.
Precursor recommendation with Magpie encoding achieves a success rate of 68\% within five attempts; it performs reasonably well because these properties reflect the material composition and generally materials with close compositions tend to be similar.
Similarly, precursor recommendation directly with the raw material composition achieves a success rate of 66\% within five attempts. 
FastText encoding \cite{kim2020inorganic} utilizes the FastText model \cite{bojanowski2017enriching} to capture information about the co-occurrences of context words around material formulas/names in the literature.
However, only 1,985 test materials can be digitized with FastText encoding due to the conflict between the limited vocabulary of n-grams and the variety of float numbers in material formulas.
The success rate using FastText encoding is 56\% within five attempts. 
Overall, the recommendation with PrecursorSelector encoding performs substantially better because Magpie and FastText encodings are more generic but not dedicated to predictive synthesis.
The PrecursorSelector encoding and MPC capture the correlation between synthesis variables and known target materials, which better extends to novel materials. 

\section*{Discussion}

Because of its heuristic nature, it is challenging to capture the decades of synthesis knowledge established in the literature. 
By establishing a materials similarity measure that is a natural handle of chemical knowledge and leveraging a large-scale dataset of precedent synthesis recipes, our similarity-based recommendation strategy mimics human synthesis design and succeeds in precursor selection.
The incorporation of precursor information into materials representations (Fig. \ref{fig:NN_framework}) leads to a quantitative similarity metric that successfully reproduces a known precursor set 82\% of the time in five attempts or less (Fig. \ref{fig:precursors_multi_attempts}).
We discuss the strengths and weaknesses of this recommendation algorithm and its generalizability to broader synthesis prediction problems.

In this work, materials similarity is learned through an automatic feature extraction process mapping a target material to the combination of precursors.
While learning the usage of precursors, useful chemical knowledge for synthesis practice is accordingly embedded in PrecursorSelector encoding. 
The first level of knowledge about materials similarity is based on composition. 
For example, to synthesize Li$_7$La$_3$Nb$_2$O$_{13}$, PrecursorSelector encoding finds Li$_5$La$_3$Nb$_2$O$_{12}$ as a reference target material (Table \ref{table:recommendation_examples}) because their difference in composition is only one Li$_2$O unit.
PrecursorSelector encoding also reflects the consideration of valence in synthesis.
Although it is not necessary to keep the valence in the precursor the same as that in the target, a precursor with similar valence states to the target is frequently used in practical synthesis \cite{he2020similarity}.
For example, to synthesize NaGa$_{4.6}$Mn$_{0.01}$Zn$_{1.69}$Si$_{5.5}$O$_{20.1}$ \cite{lv2018transition}, MnCO$_3$ was used as the Mn source because the valence state of Mn is 2+ in both the target and precursor. 
PrecursorSelector encoding finds Mn$_{0.24}$Zn$_{1.76}$SiO$_{4}$ similar to NaGa$_{4.6}$Mn$_{0.01}$Zn$_{1.69}$Si$_{5.5}$O$_{20.1}$ because the valence state of Mn is also 2+ in Mn$_{0.24}$Zn$_{1.76}$SiO$_{4}$, despite NaGa$_{4.6}$Mn$_{0.01}$Zn$_{1.69}$Si$_{5.5}$O$_{20.1}$ containing large fractions of Na and Ga while Mn$_{0.24}$Zn$_{1.76}$SiO$_{4}$ does not.
Our algorithm also captures the similarity of syntheses between compounds which have one element substituted.
For example, PrecursorSelector encoding refers to CaZnSO for synthesizing SrZnSO because the elements Ca and Sr are regarded as similar.
While such knowledge may appear obvious to the trained chemist, our approach enables it to be automatically extracted and convoluted as a vectorized representation (Fig. \ref{fig:NN_framework}), making it thereby available in a mathematical form, convenient to be used in recommendation engines or automated labs \cite{szymanski2021toward}.

Because of this customized synthesis similarity of materials and our precursor recommendation pipeline, we are able to not only recommend trivial solutions for target synthesis such as the use of common precursors, but also deal with more challenging situations.
One typical scenario is the adoption of uncommon precursors.
For example, Lal{\`e}re et al. \cite{lalere2018coupled} used NaH$_2$PO$_4$ as the source of Na and P to synthesize Na$_3$TiV(PO$_4$)$_3$, while the common precursors for Na and P are Na$_2$CO$_3$ and NH$_4$H$_2$PO$_4$, respectively.
It is not apparent to conclude from the composition of Na$_3$TiV(PO$_4$)$_3$ that the uncommon precursor NaH$_2$PO$_4$ is needed.
However, the similarity-based recommendation pipeline successfully predicts the use of NaH$_2$PO$_4$ by referring to a similar material Na$_3$V$_2$(PO$_4$)$_3$ \cite{feng20173}.
A plausible reason for the choice of NaH$_2$PO$_4$ for Na$_3$TiV(PO$_4$)$_3$ can also be inferred from the synthesis of Na$_3$V$_2$(PO$_4$)$_3$.
Feng et al. \cite{feng20173} reported that NaH$_2$PO$_4$ was used to implement a one-pot solid-state synthesis of Na$_3$V$_2$(PO$_4$)$_3$, while Fang el al. \cite{fang2015hierarchical} reported that a reductive agent and additional complex operations are needed when using Na$_2$CO$_3$ and NH$_4$H$_2$PO$_4$.
Similar outcomes may also apply to the synthesis of Na$_3$TiV(PO$_4$)$_3$.
A second example is the successful precursor recommendation for the target compound GdLu(MoO$_4$)$_3$. 
Instead of the common precursor MoO$_3$, a less common precursor (NH$_4$)$_6$Mo$_7$O$_{24}$ was adopted as the Mo source \cite{wang2018efficiently}. 
The use of (NH$_4$)$_6$Mo$_7$O$_{24}$ may facilitate the mixing of different ions in the synthesis of GdLu(MoO$_4$)$_3$. 
The adoption of uncommon precursors also provides clues in underexplored chemical spaces such as mixed-anion compounds \cite{kageyama2018expanding}. 
Taking the pentanary oxynitride material BaYSi$_2$O$_5$N \cite{yasunaga2019synthesis} as an example, the five-component system, including multiple anions, implies that many precursor combinations can potentially yield the target phase, including oxides, nitrides, carbonates, etc. 
Our recommendation pipeline correctly identifies that a combination of SiO$_2$ and Si$_3$N$_4$ facilitates the formation of BaYSi$_2$O$_5$N by referring to a quaternary oxynitride material, YSiO$_2$N \cite{kitagawa2018intense}. 
Another challenging situation is that multiple precursors may be used for the same element.
Usually, only one precursor is used for each metal/metalloid element in the target material, but exceptions do exist.
For example, CuO and CuCl$_2$ were used as the Cu source in the synthesis of Cu$_3$Yb(SeO$_3$)$_2$O$_2$Cl \cite{markina2017interplay}.
Through analogy to Cu$_4$Se$_5$O$_{12}$Cl$_2$ \cite{zhang2010synthesis}, the recommended precursor set includes both CuO and CuCl$_2$.
Moreover, it is possible to predict multiple correct precursor sets by referring to multiple similar target materials. 
For example, two different sets of precursors for LiMn$_{0.5}$Fe$_{0.5}$PO$_4$ were reported by Zhuang et al. \cite{zhuang2019synergistic} and Wang et al. \cite{wang2020synthesis}.
The recommendation pipeline predicts both by repurposing the precursor sets for LiMn$_{0.8}$Fe$_{0.2}$PO$_4$ \cite{zou2012preparation} and LiMn$_{0.9}$Fe$_{0.1}$PO$_4$ \cite{yi2011optimized}.

The recommendation of precursors presented here is still imperfect. 
The engine we present is inherently limited by the knowledge base it is trained on, thereby biasing recommendations toward what has been done previously and lacking creativity for unprecedented combinations of precursors.
For example, metals Co and Te were used in the synthesis of Li$_3$CoTeO$_6$ \cite{heymann2017li}, but no similar materials in the knowledge base use the combination of Co and Te as precursors.
Another example is that SrCO$_3$ and SrSO$_4$ were used in the synthesis of Sr$_4$Al$_6$SO$_{16}$ \cite{ndzila2020regulation}. 
Although the recommendation pipeline is, in principle, able to predict multiple precursors for the same element, a similar case using both SrCO$_3$ and SrSO$_4$ as the Sr source is not found in the knowledge base.
Both examples end up being mispredictions.
This situation could be improved when more data from text mining and high-throughput experiments \cite{szymanski2021toward} are added to the knowledge base. 
Furthermore, the success rate of the recommendation strategy may be underestimated in some cases.
For example, BaO is predicted as the Ba source for synthesizing Ca$_{7.5}$Ba$_{1.5}$Bi(VO$_{4}$)$_7$, while BaCO$_{3}$ is used in the reported synthesis \cite{dorbakov2019barium}.
Given the slight difference between BaO and BaCO$_{3}$, BaO may actually be suitable.

Besides the prediction of precursors, the similarity-based recommendation framework is a potential step toward general synthesis prediction.
The same strategy can be extended to the recommendation of more synthesis variables, such as operations, device setups, and experimental conditions, by adding corresponding prediction tasks to the downstream part of the multi-task network (Fig. \ref{fig:NN_framework}) for similarity measurement.
For example, we may infer that reduced atmosphere is necessary for synthesizing Na$_3$TiV(PO$_4$)$_3$ \cite{lalere2018coupled} because it is used in the synthesis of a similar material Na$_3$V$_2$(PO$_4$)$_3$ \cite{feng20173}. 
Moreover, synthesis constraints such as the type of synthesis method, temperature, morphology of the target material, particle size, and cost can be added as conditions of synthesis prediction.
For example, we may integrate our effort of synthesis temperature prediction to prioritize the predicted precursors within expected temperature regime. 
Our automated algorithm, mimicking human design process for the synthesis of a new target, provides a practical solution to query decades of heuristic synthesis data in recommendation engines and autonomous laboratories.

\begin{landscape}
\begin{table}[H]
	\caption{
 \textbf{Representative successful and failed examples for precursor prediction using the similarity-based recommendation pipeline in this study.}
    } 
	\centering
	\setstretch{1.35}
	\begin{tblr} {
    	colspec={ 
    	    Q[wd=0.24\linewidth, halign=l, valign=m] | 
    	    Q[wd=0.18\linewidth, halign=l, valign=m] | 
    	    Q[wd=0.29\linewidth, halign=l, valign=m] | 
    	    X[halign=l, valign=m] 
    	},
    	rowsep=0pt,
    	hline{1,Z}={2pt,solid},
        hline{2-Y}={1pt,solid},
        hspan=minimal,
	}
		Target & Reference Target(s) & Expected Precursors & Error in Recommendation   \\
		\SetCell[c=4]{l} \textit{Successful} \\
		Li$_7$La$_3$Nb$_2$O$_{13}$ \cite{peng2017synthesis} & Li$_5$La$_3$Nb$_2$O$_{12}$ \cite{van2007mechanism} & LiOH, La$_2$O$_3$, Nb$_2$O$_5$	     & N/A   \\
		NaGa$_{4.6}$Mn$_{0.01}$Zn$_{1.69}$Si$_{5.5}$O$_{20.1}$ \cite{lv2018transition} &  Mn$_{0.24}$Zn$_{1.76}$SiO$_{4}$ \cite{park2015strong}  & MnCO$_3$, Na$_2$CO$_3$, Ga$_2$O$_3$, SiO$_2$, ZnO  	     & N/A   \\
		SrZnSO \cite{chen2020creating} & CaZnSO \cite{duan2009photoluminescence} & SrCO$_3$, ZnS    & N/A   \\
		Na$_3$TiV(PO$_4$)$_3$  \cite{lalere2018coupled} & Na$_3$V$_2$(PO$_4$)$_3$ \cite{feng20173} &  NaH$_2$PO$_4$, NH$_4$VO$_3$, TiO$_2$	     & N/A   \\
        GdLu(MoO$_4$)$_3$ \cite{wang2018efficiently}  & Gd$_2$(MoO$_4$)$_3$  \cite{thirumalai2013controlled}  &  (NH$_4$)$_6$Mo$_7$O$_{24}$, Lu$_2$O$_3$, Gd$_2$O$_3$  &  N/A   \\
        BaYSi$_2$O$_5$N  \cite{yasunaga2019synthesis}   &  YSiO$_2$N  \cite{kitagawa2018intense}  	&  Si$_3$N$_4$, SiO$_2$, BaCO$_3$, Y$_2$O$_3$  &  N/A  \\
		Cu$_3$Yb(SeO$_3$)$_2$O$_2$Cl \cite{markina2017interplay} &  Cu$_4$Se$_5$O$_{12}$Cl$_2$ \cite{zhang2010synthesis}     & CuO, CuCl$_2$, SeO$_2$, Yb$_2$O$_3$       & N/A   \\
		LiMn$_{0.5}$Fe$_{0.5}$PO$_4$ \cite{zhuang2019synergistic,wang2020synthesis}  & LiMn$_{0.8}$Fe$_{0.2}$PO$_4$ \cite{zou2012preparation}, LiMn$_{0.9}$Fe$_{0.1}$PO$_4$ \cite{yi2011optimized} & MnCO$_3$, FeC$_2$O$_4$, LiH$_2$PO$_4$; Mn(CH$_3$COO)$_2$, FeC$_2$O$_4$, LiH$_2$PO$_4$    & N/A   \\
        \SetCell[c=4]{l} \textit{Failed} \\
		Li$_3$CoTeO$_6$ \cite{heymann2017li} & LiCoO$_2$ \cite{alcantara1999x} & Co, Te, Li$_2$CO$_3$  & Co$_3$O$_4$, TeO$_2$, LiOH   \\
        Sr$_4$Al$_6$SO$_{16}$ \cite{ndzila2020regulation} & SrAl$_2$O$_4$ \cite{zhu2009encapsulation} & SrCO$_3$, SrSO$_4$, Al(OH)$_3$ & SrCO$_3$, H$_2$SO$_4$, Al(OH)$_3$  \\
        Ca$_{7.5}$Ba$_{1.5}$Bi(VO$_{4}$)$_7$ \cite{dorbakov2019barium} & Bi$_3$Ca$_9$V$_{11}$O$_{41}$ \cite{radosavljevic2000synthesis} & BaCO$_{3}$, NH$_4$VO$_3$, CaCO$_3$, Bi$_2$O$_3$ & BaO, NH$_4$VO$_3$, CaCO$_3$, Bi$_2$O$_3$ \\
	\end{tblr}
	\label{table:recommendation_examples}
\end{table}
\end{landscape}

\section*{Materials and Methods}

\subsection*{Representation learning for similarity of materials} \label{sec:detailed_NN}

The neural network consists of an encoder part for encoding target materials and a task part for predicting variables related to precursor selection. 
The encoder part $f$ is a three-layer fully connected submodel transforming the composition of the target material $\bm{x}$ into a 32-dimensional target vector $\bm{u}=f(\bm{x})$.
The input composition is an array with 83 units showing the fraction of each element.
The reduced dimension of the encoded target vector is inspired by the bottleneck architecture of autoencoders \cite{bengio2013representation}. 
By limiting the dimension of the encoded vector, the network is forced to learn a more compact and efficient representation of the input data, which is more appropriate for the precursor selection-related downstream tasks \cite{tschannen2018recent}. 
The task part uses different network architectures for different tasks of prediction, including precursor completion and composition recovery in this work. 
The masked precursor completion (MPC) task replaces part of the precursors with a placeholder ``[MASK]'' \cite{devlin2018bert} at random and uses the remaining precursors as a condition to predict the complete precursor set for the target material, which is formulated as a multi-label classification problem \cite{herrera2016multilabel}. 
An attention block $g_{proj}$ \cite{vaswani2017attention} is used to aggregate the target vector and the vectors for conditional precursors as a projected vector $\bm{v}=g_{proj}(\bm{u};\bm{p}_1,\bm{p}_2,\dots)$ with dimensionality of 32.
Then, $\bm{v}$ is passed to the precursor classification layer represented by a $ 417 \times 32 $ matrix $\bm{P}$, of which each row is the 32-dimensional vector representation of a potentially used precursor $\bm{p}_i$. 
To avoid having too many neural network weights to learn, the precursor completion task only considers 417 precursors used in at least five reactions in the knowledge base.
The probability to use each precursor is indicated by $\operatorname{sigmoid}(\bm{p}_i^{\top}\bm{v})$, allowing non-exclusive prediction of multiple precursors \cite{herrera2016multilabel}. 
Here, $\bm{v}$ acts as a probe corresponding to the target material projected in the precursor space and is used to search for $\bm{p}_i$'s with similar vector representations via a dot product.
The conditional precursors input to $g_{proj}$ share the same trainable vector representations as $\bm{p}_i$'s.
Circle loss \cite{sun2020circle} is used because of its benefits in capturing the dependency between different labels in multi-label classification and deep feature learning.
The composition recovery task is a two-layer fully connected submodel decoding back to the chemical composition $\bm{x}$ from the target vector $\bm{u}$, similar to the mechanism of autoencoders \cite{bengio2013representation,hinton2006reducing}.  
Mean squared error loss is used because it is the most popular for regression.
More tasks predicting other synthesis variables such as operations and conditions can be appended in a similar fashion.
To combine the loss functions in this multi-task neural network, an adaptive loss \cite{kendall2018multi} is used to automatically weigh different loss by considering the homoscedastic uncertainty of each task.

\subsection*{Baseline models} \label{sec:baselines}

\paragraph*{``Most frequent''.} This baseline model ranks precursor sets based on an empirical joint probability without considering the dependency of precursors (Fig. \ref{fig:precursor_statistics}B).
Assuming that the choices of precursors are independent from each other, the joint probability of selecting a specific set of precursors can be estimated as the product of their marginal probabilities.  
For each metal/metalloid element, different precursors can be used as the source. 
The marginal probability to use a precursor is estimated as the relative frequency of using that precursor over all precursors contributing the same metal/metalloid element.
For example, the precursor set ranked in first place is always the combination of common precursors for each metal/metalloid element in the target material, which is also typically the first attempt in the lab.  

\paragraph*{``Magpie encoding''.} This baseline model uses the same recommendation strategy as Fig. \ref{fig:pipeline}, except that the similarity is calculated using Magpie encoding \cite{ward2016general,ward2018matminer}.
The composition of each target material is converted into a vector consisting of 132 statistical quantities such as the average and standard deviation of various elemental properties.
The cosine similarity is used, as shown in Eq. \ref{eq:similarity_encoding}.
When the precursors from the reference target material cannot cover all elements of the novel target, the common precursors for the missing elements are supplemented because MPC (Fig. \ref{fig:NN_framework}B) is only trained for PrecursorSelector encoding.

\paragraph*{``FastText encoding''.} Similar to the baseline of ``Magpie encoding'', this baseline model uses the same recommendation strategy as Fig. \ref{fig:pipeline}, except that the similarity is calculated using FastText encoding \cite{kim2020inorganic}.
The formula of each target material is converted into a 100-dimensional vector using the FastText model trained with materials science papers \cite{kim2020inorganic}.
The total number of target materials tested in this baseline model is 1,985 instead of 2,654 because some n-grams such as certain float numbers corresponding to the amount of elements are not in the vocabulary.

\paragraph*{``Raw composition''.} Similar to the baseline of ``Magpie encoding'', this baseline model uses the same recommendation strategy as Fig. 1, except that the similarity is calculated using the cosine similarity of raw material composition. 
The formula of each target material is converted into an 83-dimensional vector corresponding to the fraction of each element. 

\subsection*{Data preparation} \label{sec:data_preparation}

In total, 33,343 inorganic solid-state synthesis recipes extracted from 24,304 materials science papers \cite{kononova2019text} were used in this work.
Because some material strings (e.g., $ \operatorname{Ba}_{1-x} \operatorname{Sr}_{x} \operatorname{TiO}_3 $) extracted from the literature contain variables corresponding to different amounts of elements, we substituted these variables with their values from the text to ensure that a material in any reaction only corresponds to one composition, resulting in 49,924 expanded reactions and 28,598 target materials.
An ideal test for generalizability and applicability of this method would be to synthesize many entirely new materials using recommended precursors. 
In the absence of performing extensive new synthesis experiments, we designed a robust test to simulate precursor recommendation for target materials that are new to the trained model. 
We split the data based on the year of publication, i.e., training set (or knowledge base) for reactions published by 2014, validation set for reactions in 2015 and 2016, and test set for reactions from 2017 to 2020.
In addition, to avoid data leakage where the synthesis of the same material can be reported again in a more recent year, we placed reactions for target materials with the same prototype formula in the same data set as the earliest record.
The prototype formula was defined as the formula corresponding to a family of materials including (1) the formula itself, (2) formulas derived from a small amount ($<0.3$) of substitution (e.g., $ \operatorname{Ca_{0.2}La_{0.8}MnO_3} $ for prototype formula $ \operatorname{LaMnO_3} $), and (3) formulas able to be coarse-grained by rounding the amount of elements to one decimal place (e.g., $ \operatorname{Ba_{1.001}La_{0.004}TiO_3} $ for the prototype formula $ \operatorname{BaTiO_3} $).
In the end, the number of reactions in the training/validation/test set was 44,736/2,254/2,934 from 29,900/1,451/1,992 original recipes.
The number of target materials in the training/validation/test set was 24,304/1,910/2,654, respectively.

\nocite{liu2019roberta,kingma2014adam,prechelt2002early}


\bibliography{sciadvbib}
\bibliographystyle{ScienceAdvances}

\subsection*{Acknowledgements}
%
The authors thank Prof. Wenhao Sun, Dr. Anubhav Jain, Prof. Elsa Olivetti, and Dr. Olga Kononova for valuable discussions.

\paragraph*{Funding:} The U.S. Department of Energy, Office of Science, Office of Basic Energy Sciences, Materials Sciences and Engineering Division (DE-AC02-05-CH11231, D2S2 program KCD2S2).
The Assistant Secretary of Energy Efficiency and Renewable Energy, Vehicle Technologies Office, U.S. Department of Energy (DE-AC02-05CH11231).
The National Science Foundation (DMR-1922372).
Savio computational cluster resource provided by the Berkeley Research Computing program at the University of California, Berkeley (supported by the UC Berkeley Chancellor, Vice Chancellor for Research, and Chief Information Officer). 

\paragraph*{Author Contributions:} Conceptualization: TH, GC.
Methodology: TH, HH, CJB, ZW, KC.
Investigation: TH.
Visualization: TH.
Supervision: GC.
Writing—original draft: TH, GC.
Writing—review \& editing: TH, HH, CJB, ZW, KC, GC.

\paragraph*{Competing Interests:} The authors declare that they have no competing interests.

\paragraph*{Data and materials availability:} 
All data needed to evaluate the conclusions in the paper are present in the paper and/or the Supplementary Materials. 
The code for the similarity-based synthesis recommendation algorithm and the data supporting the findings of this study are available at the Dryad repository https://doi.org/10.6078/D1XD96 and the Github repository 
https://github.com/CederGroupHub/SynthesisSimilarity.

\end{document}